\documentclass{article}

\usepackage{arxiv}

\usepackage[utf8]{inputenc} 
\usepackage[T1]{fontenc}    
\usepackage{hyperref}       
\usepackage{url}            
\usepackage{booktabs}       
\usepackage{amsfonts}       
\usepackage{nicefrac}       
\usepackage{microtype}      
\usepackage{lipsum}
\usepackage{graphicx}
\graphicspath{ {./images/} }

\title{Queueing Systems with Some Versions of Limited Processor
 Sharing Discipline 
}

\author{
 Alencar M.S. \\
Department of Electrical Engineering\\
Federal University of Campina Grande\\
Campina Grande, Paraíba, Brazil\\ 
  \texttt{malencar@iecom.org.br } \\
 \And
 Tatashev A.G. \\
  Department of Higher Mathematics\\
  Moscow Automobile and Road Construction\\
  State Technical University (MADI) \\
  Moscow, Leningradsky avenue, 64, Russia  \\
  \texttt{a-tatashev@yandex.ru} \\
   \And
Seleznjev O.V.\\
Umea University, Sweden\\
  \texttt{oleg.seleznjev@umu.se} \\
   \And
 Yashina M.V. \\
  Department of Higher Mathematics\\
  Moscow Automobile and Road Construction\\
  State Technical University (MADI) \\
  Moscow, Leningradsky avenue, 64, Russia  \\
  \texttt{mv.yashina@madi.ru} \\
}

\begin{document}
\maketitle
\begin{abstract}
The paper considers a queueing 
system with limited processor sharing. No more than $n$ jobs may be served simultaneously.  This system may be used for modeling bandwidth sharing in wireless communication systems and processes of service in computer networks.  If there are $n$ jobs in the considered queueing system and a new job arrives, then the arriving job is lost or the service of a job is interrupted and this job is lost. We study two rules to choose the job to be lost. In accordance with one of these rules, the job with the shortest remaining length is lost. Relations are obtained between the state probabilities of considered system and the state probabilities of the corresponding unlimited processor sharing system. These relations allows to compute the state probabilities for considered system if the state probabilities for the unlimited processor sharing system are known. 

In the case of Poisson arrival process, the probability that the server capacity is exhausted is equal to the probability that a job is lost. 
We have obtained an explicit formulas for the stationary state probabilities and the loss probability for this case.  These probabilities are invariant under the job length distribution under the condition that the average value of the length is fixed.   
The second considered rule is the First Come, First Displaced (FCFD) rule. If the service time is equal to a prescribed constant with probability 1, then the shortest remaining length job loss discipline is equivalent to the FCFD discipline. In the case of arbitrary length distribution, the loss probability for the shortest remaining length job loss discipline is greater than the loss probability for another discipline, and, in particular, the FCFD discipline. Hence we have an upper bound for the loss probability in the system with an arbitrary rule for choosing a job to be lost.
 We have found that, in the case of FCFD rule and constant service length or in the case of the shortest remaining length job loss, the stationary lost probability in a limited egalitarian processor sharing system can be not monotonically decreasing function of the maximum admissible number of jobs in the system for a fixed arrival rate.

\end{abstract}


\section{Introduction}


Processor sharing disciplines are widespread in communications system and computer networks. The basic principle to organize requests for the processor time is that jobs are alternately served during small intervals (slots) of time,~\cite{Yashkov} .
 The implementation of this principle to organize a computing process is carried out using a scheduling algorithm. The principle of service quantization is a way to reduce the response time of the 
system to short jobs due to the delay in processing long jobs in the absence of information about the remaining time of jobs. Processor sharing is efficient in the sense of delay minimization if the service time variation is higher than the variation of the exponential service time distribution.    

In the case of egalitarian processor sharing, all jobs contained in the system at present time are served with the same rate. 
For the systems with unlimited egalitarian processor sharing, the stationary state distribution for the M/G/1 queueing system is the same as in the M/M/1 queueing system with the 
First Input, First Output (FIFO) discipline ~\cite{ Kleinrock}, ~\cite{Chandy}. In a particular case in that the total service rate is constant, one may suppose that each job is served with rate $1/i$ if $i$ jobs are in the system.

In versions of this discipline, a certain minimum share of the bandwidth of the computer or channel has to be guaranteed to provide acceptable quality of service to a customer. Therefore the processor sharing with a finite capacity or limited processor sharing is often used. 
In  ~\cite{Telek},  ~\cite{ Katchner} , a queueing system with the limited processor sharing was studied. In~\cite{Telek}, it is assumed that a job arriving when the capacity of the server is exhausted joins the buffer to obtain service later, and, in~\cite{ Dudin}, it is assumed that such a job is lost. 
It is noted in~\cite{ Dudin} that it seems that the model with customer loss better suits, e.g., for modeling bandwidth sharing in wireless communication networks.  In~\cite{Telek}, it is assumed that the service rate of a job is a prescribed value depending on the number of simultaneously served jobs. In~\cite{ Dudin} , it is assumed that the rate of a job is inversely proportional to the number of customer in service. In  ~\cite{Dudin}, possible impatience of customers is accounted, i.e., a job may be impatient and leave the system before service completion due to long processing. It assumed in~\cite{Telek},  ~\cite{Dudin} that jobs arrive in accordance with the Markovian arrival process and the service time has a phase type distribution.       

In accordance with First Come, First Displaced (FCFD) protocol ~\cite{ Katchner}, the job arrived first is interrupted. It is mentioned in~\cite{ Katchner} that the FCFD discipline is useful when the importance of call decreases with the elapsing time.  In~\cite{ Katchner} a loss system was considered, in which jobs of higher priorities preempts the service of lower priorities.

 The paper~\cite{Tatashev91} considered a queueing system with the loss of the shortest remaining loss job. Let $n$ be the number of servers in the system. It is proved in~\cite{Tatashev91} that the probability that all servers are busy is equal to the probability that, in the corresponding system with infinite number of servers, there are no less than $n$ jobs. In the case of a Poisson arrival process, 
the loss probability for the system considered in~\cite{Tatashev91} is an upper bound for the loss probability in the system with an arbitrary rule of choosing job to be lost. In the case of G/D/n/0 system, the shortest remaining length job discipline loss is equivalent to the loss FCFD discipline. In the case of 
a Poisson arrival process, the stationary distribution for number of jobs in the shortest remaining loss discipline system does not 
depend on the service time distribution under the assumption that the expectation of the service time is prescribed. A class of disciplines 
invariant in the sense that the stationary state probabilities does not       
depend of the service time distribution type consists of disciplines conserving Poisson process, i.e., the output of systems with these disciplines is described with a Poisson process if the system input is described with the Poisson process ~\cite{Chandy}. For systems with these disciplines, the stationary state probabilities are the same as for the corresponding queueing system with First Input, First Output discipline and the exponential distribution of job service. If the nodes of a queueing network are queueing systems of these class, then the stationary network state probabilities have product form~\cite{Chandy}.  It is well known that the M/G/n/0 system with loss of arriving jobs, the unlimited egalitarian processor sharing M/G/1 system, and the M/G/$\infty$ system are invariant systems with Poisson output. In the case of M/G/n/0, the output process is formed by all served and lost jobs. In ~\cite{Pechinkin79}, an invariant system has been found, which does not conserve the Poisson process.  In~\cite{Pechinkin79}, the system M/G/1/k with the discipline such that first the job with the shortest remaining service time (length) is served. If the length of arriving job is less than the remaining length of served job, then the arriving job is served from the arrival moment, and the that was served is interrupted. The maximum number of jobs served simultaneously (the buffer capacity) equals $k.$ The interrupted job is served from the point of interruption. If $k=1,$ i.e., the buffer capacity is equal to 1, then the stationary probabilities of system states do not depend on the service time distribution under the condition that the average value of the service time is prescribed. If $k>1,$ the stationary distribution for number of jobs being in the system is not invariant, i.e., this distribution depends on the service time distribution. In   ~\cite{Pechinkin83}, a queueing system was considered such that the discipline differs from the shortest remaining length job service discipline in the following. Let a job arrives at time $t_0$ and  Let a job arrive at time $t_0$ and the length of this job is greater than remaining length of the served demand. Then the job arriving at time $t_0$ occupies the first place of the buffer, i.e., the service of this job will begin earlier than the service of any job being in the system at time $t_0.$ The distribution for the number of jobs being in the system do not depend on the distribution of the service time distribution in the case of any finite capacity of the buffer and in the case of infinite buffer under the average value of the service time is prescribed. This system does not belong to the class of invariant discipline conserving a Poisson process at the output. The form of the stationary state probabilities for  the  corresponding network is not multiplicative. The stationary state probabilities are not the same as for the system M/M/1/k with the FIFO discipline. However these stationary state probabilities are equal to the system M/D/1/k with the FIFO discipline since, in the case of  constant service time, the discipline, considered in  ~\cite{Pechinkin83}, and the FIFO discipline are equivalent to each other. In papers,~\cite{Pechinkin79}, ~\cite{Pechinkin83} queueing systems are found with invariant state probabilities are found such that these systems do not conserve the Poisson process. It is proved in ~\cite{Shrage} that the shortest remaining time job service discipline minimizes the average sojourn time for the class of job conserving disciplines. Hence the value of the average sojourn time for the M/D/1 system with FIFO discipline is an upper bound for the average sojourn time for the M/G/1 system with the shortest remaining time job service discipline.

In the paper  ~\cite{Pechinkin83}, the following approach was developed to study the system considered in ~\cite{Pechinkin83}, and this approach was used to study the characteristics of queueing systems, considered, e.g., in~\cite{Tatashev95}, ~\cite{Pechinkin99}, ~\cite{Ivnitskii},~\cite{Tatashev03}.      
Most of these systems are versions of the Last Input, First Output (LIFO) discipline. In accordance with this approach, a set of states in the state space is considered. A stochastic process related to the system state space is considered for the time intervals such that, on these intervals, the process belongs to the distinguished set of states. The conditional stationary distribution of the original process under the condition that the values of the process belong to the distinguished set is the same as the distribution of the process for the distinguished time intervals. Some systems studied with aid of this approach are invariant in the considered sense (for example,  the system studied in~\cite{Tatashev91}, ~\cite{Pechinkin83} are invariant), and some systems are not invariant (for example,  the system studied in~\cite{Pechinkin99},  ~\cite{Tatashev03}  are not invariant). For the system considered in~\cite{Ivnitskii}, there are both invariant and non-invariant versions. 

The paper~\cite{Tatashev95} considered a queueing system with a batch Poisson arrival process and preempted-resume, preempted-restart, preempted-repeat-identical, and preempted-loss LIFO disciplines.     

In  ~\cite{ATYa20} an approximate approach was developed to compute the average sojourn time for the limited processor sharing system with FIFO choosing job from the buffer.
This queueing system was interpreted as a mathematical model used to evaluate characteristics of the mobile connection between drivers moving on a highway, or between drivers and a facility located on the same highway.  

This paper consider a loss queueing system with limited processor sharing. Section~2 studies the system with the shortest remaining length job loss.
Relations are obtained between the state probabilities of considered system and the state probabilities for the corresponding unlimited processor sharing system. These relations allow to compute the state probabilities for considered system if the state probabilities for the unlimited processor sharing system is known. In the case of Poisson arrival process, the probability that the server capacity is exhausted
is equal to the probability of the job loss. We have obtained explicit formulas for the stationary state probabilities and the loss probabilities for this case. 
We have found that the, in the case of FCFD rule and constant service length or the shortest remaining length job rule, 
the stationary lost probability in a limited egalitarian processor sharing system can be not monotonically decreasing function for 
a fixed arrival rate unlike the system with the loss of arriving jobs, for which the loss probability is a decreasing function.    
If the service time is equal to a prescribed constant with probability 1, then the shortest remaining length job loss discipline is equivalent to the FCFD discipline. In the case of arbitrary distribution of job length, the loss probability for the shortest remaining length job loss discipline is greater than the loss probability for any other discipline, and, in particular, the FCFD discipline. Hence we have an upper bound for the loss probability in the system with an arbitrary rule for choosing a job to be lost. The system with of the FCFD discipline is considered in Section~3. We have obtained an explicit formula for the average sojourn time  the case in  
that, with prescribed probabilities, the job length is distributed exponentially or this length is equal to 0.

\section{
 Limited processor sharing with the shortest remaining length 
job loss rule
}
\label{section:LPS}

The paper~\cite{Tatashev91} considered a multi-server queueing system with arbitrary arrival process and arbitrary service time distribution The number of servers is equal to $n.$ If there are $n$ jobs in the system, and a new job arrives, then the job with the minimum remaining service time is lost.

The loss probability for this system may be regarded as an upper bound for the loss probability in the G/G/n/0 queueing system with an arbitrary
rule for choosing a job to be lost.        

Let us consider the G/G/$\infty$ system with the same arrival process and the same length distribution as for the considered G/G/n/0 system.

Suppose the initial state of these two systems is the same. It is proved in ~\cite{Tatashev91} that the probability that a job arriving to the loss system at time $t_0$  
is lost equals the probability that, at time $t_0-0,$ there are more than $n-1$ jobs in the G/G/$\infty$ system.       

For the Poisson arrival process with rate $\lambda$ and the average service time $b$ we have~\cite{Tatashev91}]
$$
p_i=\left\{
\begin{array}{l}
e^{-\lambda b}\frac{(\lambda b)^i}{i!},\  i=0,1,\dots,n-1,\\
1-e^{-\lambda b}\sum\limits_{i=0}^{n-1}\frac{(\lambda b)^i}{i!},\ i=n,\\
\end{array}
\right.
$$ 
where $p_i$ is the stationary probability that in the loss system there are $i$ jobs, $i=0,1,\dots,n,$ and the stationary loss probability is equal to $p_n.$

Let us consider a limited processor sharing system with arbitrary arrival process and arbitrary job length distribution. There are no more than $n$ jobs in the system at any time.  If there are $i\le n$ jobs in the system, then each job is served with the rate $c_i,$ $i=1,\dots,n.$ If $n$ jobs are served, a job arrives, and the length of this job is less than the minimum remaining length of any served job, then the arriving job is lost. If the length of the arriving job is not less than the minimum remaining length of the served jobs, then the job with the minimum remaining length is interrupted and lost, and the arriving job is served. Let $c(t)$ be the service rate at time $t.$ We assume that, if a job is not served at time $t,$ then $c(t)=0.$ Let a job arrive at time $t_1.$ The service of the job ends at time $t_2$ such that 
$$\int\limits_{t_1}^{t_2}c(t)\,dt=l,$$
where $l$ is the job length. This system is called the system~$S_1.$ 

Let us introduce an unlimited processor sharing system with the same arrival process and the same distribution of job length as in the system $S_1.$ If there are $i\le n$ jobs in the system, then each job is served with rate $c_i,$ 
$i=1,\dots,n.$ If there are $i>n$ jobs in the system, then each job is served with rate $c_n.$ This system is called the system~$S_2.$
          
Denote by $X(t)$ and $\hat{X}(t)$ the number of the jobs at time $t$ in the systems $S_1$ and $S_2$ respectivly. 
Denote by $p_i(t)$ the probability that $X(t)=i,$ $i=0,1,2,\dots,n.$ Denote by $p_i(t)$ the probability that $X(t)=i,$ $i=0,1,2,\dots,n.$ 

Let us generalize the results obtained in ~\cite{Tatashev91}. 
\vskip 3pt
{\bf Theorem 1.} {\it The following equalities hold:
$$
p_i(t)=\left\{
\begin{array}{l}
\hat{p}_i(t),\  i=0,1,\dots,n-1,\\
1-\sum\limits_{i=0}^{n-1}\hat{p}_i(t).\\
\end{array}
\right.
$$ 
\vskip 3pt 
Proof.} Let us consider the system $S_1$ under the assumption that there is an infinite set of supplementary positions. The jobs lost in the system $S_1$ are directed to the supplementary positions for the service or the completion of the service. Obviously, a supplementary position may be busy only when there are $n$ jobs in the system $S_1$
(on the main positions). During the time of the job service on a supplementary position,
the number of jobs in the system continues to be equal to $n$ because the remaining length of any job served in the system $S_1$ is greater than the remaining length of any job served on a supplementary position. Thus, if there are busy supplementary positions, there are $n$ jobs in the system. To complete the proof, it is sufficiently to take into account that the set consisting of positions of the system $S_1$ and the supplementary       
positions is equivalent to the system $S_2.$ 
\vskip 3pt
Suppose the limits ({\it stationary state probabilities})
$$p_i=\lim \limits_{t\to \infty}p_i(t),$$
$$\hat{p}_i=\lim\limits_{t\to \infty}\hat{p}_i(t)$$
exist and do not depend on the initial state. 
\vskip 3pt
{\bf Corollary 1.} {\it 
Then the following equalities hold
$$
p_i=\left\{
\begin{array}{l}
\hat{p}_i,\  i=0,1,\dots,n-1,\\
1-\sum\limits_{i=0}^{n-1}\hat{p}_i.\\
\end{array}
\right.
$$ 
}
\vskip 3pt 
This statement follows from Theorem~1. 
\vskip 3pt 
{\bf Theorem 2.} {\it Suppose that the input of the system
$S_1$ is a Poisson process with rate depending  
on the number of jobs in the system, and $\lambda_i$ is the rate of process when there are $i$ jobs in the system, $b$ is the expectation of job length. Then we have  
$$
p_i=\left\{
\begin{array}{l}
p_0\prod\limits_{j=1}^i\rho_j,\  i=0,1,\dots,n-1,\\
p_0\frac{n!\prod\limits_{j=1}^n\rho_j}{n^n\rho^n}\left(e^{n\rho}-
\sum\limits_{i=0}^{n-1}\frac{(n\rho)^i}{i!}\right),\ i=n,
\end{array}
\right.
$$ 
where
$$\rho_i=\frac{\lambda_i b}{ic_i},\ i=1,\dots,n,$$
$$\rho=\rho_n,$$
$$p_0=\left(1+\sum_{i=1}^{n-1}\prod_{j=1}^i\rho_j+
\frac{n!\prod\limits_{j=1}^n\rho_j}{n^n\rho^n}
\left(e^{n\rho}-\sum\limits_{i=0}^{n-1}\frac{(n\rho)^i}{i!}
\right)\right)^{-1}.\eqno(1)
$$
\vskip 3pt
Proof.} For the stationary state probabilities of the unlimited processor sharing system $S_2,$ we have ~\cite{Chandy}
$$p_i=p_0\prod\limits_{j=1}^i\rho_j,\ i=1,\dots,n,$$
$$p_i=p_0\frac{(n\rho)^{i-n}}{\prod\limits_{j=n+1}^i j }\prod\limits_{j=1}^n\rho_i,\ i=n+1,n+2,\dots,$$ 		
$$p_0=\left(1+\sum\limits_{i=1}^n\prod\limits_{j=1}^i\rho_j+
\sum\limits_{i=n+1}^{\infty}\frac{(n\rho)^{i-n}}{\prod\limits_{j=n+1}^i j }\prod\limits_{j=1}^n\rho_i\right)^{-1}.\eqno(2)$$
The formula (2) may be represented in the form (1). From this
and Theorem~1, we get Theorem~2. 
\vskip 3pt
In accordance with well known fact~\cite{ATYa21}, the loss probability for a loss queueing system with a Poisson arrival process with a constant rate $\lambda$ equals the  probability that all servers are busy. Therefore, the stationary loss probability equals $p_n.$
\vskip 3pt
{\bf Corollary 2.} {\it If 
$$\lambda_0=\lambda_1=\dots=\lambda_n,$$
$$c_i=\frac{1}{i},\ i=1,\dots,n,$$
then the stationary loss probability is equal to
$$p_n=\frac{e^{n\lambda_n b}-\sum\limits_{i=0}^{n-1}\frac{(n\lambda b)^i}{i!}}
{\frac{n^n}{n!}\sum\limits_{i=0}^{n-1}(\lambda b)^i+
e^{n\lambda b}-\sum\limits_{i=0}^{n-1}\frac{(n\lambda b)^i}{i!}}.\eqno(3) 
$$
}
\vskip 3pt
{\bf Corollary 3.} {\it If jobs arrive in accordance with Poisson process, then the stationary state probabilities and the stationary loss probability do not depend on the job length distribution under the assumption that the expectation of the job length is prescribed.     
}  
\vskip 3pt
We say that the system $S_1$ is in the state $E_i^1$ if there are $i$ jobs in the system $i=1,\dots,n.$ We say that the system $S_2$ is in the state $E_i^2$ $(i=0,1,\dots,n-1)$ if there are $i$ jobs in this system, and we say that the system $S_2$ is in the state $E_n^2$ if there are no less than $n$ jobs in the system.     
\vskip 3pt
{\bf Theorem 3.} {\it For the states of the systems $S_1,$ $S_2,$ the following equality holds:
$$P^1(t_1,\dots,t_m;i_1,\dots,i_m)=P^2(t_1,\dots,t_m;i_1,\dots, i_m),$$ 
where $P^j(t_1,\dots,t_m;i_1,\dots,i_m)$ is the probability that, at time $t_1,$ the system $S_{i_1}^j$ $(j=1,2)$ is in the  state $E_{i_1}^j;$ at time $t_2,$ the system $S_j$  is in the  state $E_{i_2}^j,$ $j=1,2,$ etc. 
}
\vskip 3pt
The proof of Theorem 3 is analogous to the proof of Theorem 2. 
\vskip 3pt
{\bf Theorem 4.} {\it 
The stationary distribution of the time interval when there no jobs for the system $S_1$ (the stationary distribution of an idle period) is the same as for the system $S_2.$}   
\vskip 3pt  
The proof of Theorem 4 is also analogous to the proof of Theorem 2. 
\vskip 3pt 
{\bf Remark 1.} We may compute the average sojourn time of job in the system $S_1$ using the following approach. We compute the stationary state probabilities 
$p_0,p_1,\dots,p_n$ using Theorem~1. The average number of jobs in the system is equal to
$$L=\sum\limits_{i=1}^np_ii.$$      
Using Little's law, we get $V=L/\lambda,$ where $V$ is the average sojourn time.

\section{
Limited processor sharing loss system with FCFD rule
}
\label{section:LPSFCFDR}
\subsection{
 System with constant job length}

Consider a queueing system, which differs from the system considered in Section~2 in that, if there are $n$ jobs in the system and a new job arrives, then the arriving job interrupts the service of the job that arrived first.

Suppose the length of job is equal to $b$ with the probability~1. Then the system is equivalent to the corresponding system with the discipline considered 
in Section~\ref{section:LPS}, and the statement analogous to Theorem~1 holds for the system, considered in this subsection.

\vskip 3pt 
{\bf Remark 2.} We may compute the average sojourn time of job in the system using Remark~1. We can compute the average sojourn time of a job under the condition that the job is lost using Remark~1 and the fact that the service time is equal to prescribed constant under the condition that the job is lost.       
\vskip 3pt 
{\bf Example 1.} Suppose that $c_1=\dots=c_n=1,$ the jobs arrive in accordance with a Poisson process with rate $\lambda,$ and the length of any job
is equal to $b.$ Then the stationary loss probability is equal to
$$p_n=1-\sum\limits_{i=0}^{n-1}e^{-\lambda b}\frac{(\lambda b)^i}{i!}.\eqno(4)$$   
We may get (4) taking into account that, in the considered case, the loss probability is equal to the probability that a job is lost if and only if less than $n$ jobs arrive during the service time of the job.

\subsection{
 System with Poisson arriving process and job length distribution $B(x)=1-\alpha e^{-\mu x}$}

Let us consider the system differ from the system considered in Subsection~3.1 in that the job length has the distribution 
$$B(x)=1-\alpha e^{-\mu x},\ x\ge 0.\eqno(5)$$  
Then we have for the average job service time
$$b=\frac{\alpha}{\mu}.$$
 Assume that the jobs  arrival in accordance with a Poisson process, and the arrival rate equals $\lambda_i$ if there are $i$ jobs in the system, $i=0,1,\dots,n.$ This system is called the system $S_1.$ 
   
\vskip 3pt
{\bf Theorem 5.} {\it If (4) holds for the job length distribution, then, for the system $S_1$ the stationary state probabilities
may be computed as follows   
$$p_i=p_0\prod_{j=1}^i \rho_j,\ i=1,\dots,n,\eqno(6)$$  
$$p_0=\left(1+\sum\limits_{i=1}^n\prod_{j=1}^i\rho_j\right)^{-1},\eqno(7)$$
$$\rho_i=\frac{\alpha \lambda_i}{ic_i\mu}, i=1,\dots,n-1,\eqno(8)$$  
$$\rho_n=\frac{\alpha \lambda_n}{(1-\alpha)\lambda_n+nc_n\mu}.\eqno(9)$$ 
The stationary loss probability is equal to $p_n.$
\vskip 3pt
Proof.} 
Let us introduce the system $S_2.$ This system differs from the system $S_1$ in that the arrival rate equals $\alpha\lambda_i$ if there are $i$ jobs in the system, and the  
length of job is distributed exponentially with the average value $\mu^{-1}.$ Denote by $p_i,$ $\hat{p}_i$ the steady probability that there are $i$ jobs in the systems $S_1$ and $S_2$ respectively, $i=0,1,\dots,n-1.$ 

Obviously, $p_i/p_0=\hat{p}_i/\hat{p}_0,$ $i=1,\dots,n-1,$     
$$p_i=p_0\prod_{j=1}^i \rho_j,$$  
$$\rho_i=\frac{\alpha \lambda_i}{ic_i\mu}, i=1,\dots,n-1,$$  
$$p_0=\left(1+\sum\limits_{i=1}^n\prod_{j=1}^i\rho_j\right)^{-1}.$$
    
The time interval in that there are $n$ jobs in the system $S_1$ ends when the service of a job ends or a job of zero length arrives. Therefore the average duration of
this time interval equals $1/((1-\alpha)\lambda_n+nc_n\mu).$ Therefore,
$$p_n= p_{n-1}\rho_n,$$   
where $\rho_n$ is computed in accordance with (9). The 
stationary loss probability equals.
$$p_n=p_0\prod_{i=1}^n\rho_i.\eqno(10)$$
Theorem 5 has been proved.
\vskip 3pt
Denote by $\beta(s)$ is the Laplace--Stieltjes transform for a 
random value such that, with the probability $1-\alpha,$ this value is equal to~0, and, with the probability $\alpha,$ this random value is distributed exponentially with the expectation 
equal to $(nc_n\mu)^{-1}.$  We have
$$\beta(s)=1-\alpha+
\alpha\cdot \frac{nc_n\mu}{s+nc_n\mu},$$
$$\beta(s)=1-\frac{\alpha s}{s+nc_n\mu},$$   
$$\frac{1-\beta(s)}{\beta(s)}=\frac{\alpha s}
{(1-\alpha)s+nc_n\mu}.\eqno(11)$$  
 Combining (9) and (11), we get
 $$\rho_n=\frac{1-\beta(\lambda_n)}{\beta(\lambda_n)}.\eqno(12)$$ 
Thus we can compute the value $p_n,$ which is equal to the loss probability computed in accordance with  (6)--(8), (12).   
  
Suppose $n=1.$ Then (6)--(8), (12) hold for any job length distribution ~\cite{Tatashev95}.   
 
\vskip 3pt
{\bf Remark 3.} It follows from (6)--(10) that  
$$\lim\limits_{\mu\to 0}\rho_i=\infty,\ i=1,\dots,n-1,$$ 
$$\lim\limits_{\mu\to 0}\rho_n=\frac{\alpha}{1-\alpha},$$
and we get, for the loss probability $p_n,$ 
$$\lim\limits_{\mu\to 0}p_n=\frac{\alpha}{(1-\alpha)+\alpha}=\alpha.$$

\subsection{
 Results of computations}

In Table 1, we represent the results of computing the values of loss probabilities. Suppose $c_1=\dots=c_n=1,$ $b=1.$ Denote by $\hat{p}_n$ the value 
computed in accordance with (3). Denote by $p_n^*$ the value computed in accordance with Erlang's loss formula. This value equals the stationary loss 
probability for the system with loss of arriving job. In the case of FCFD loss rule, the value $p_n^*$ is equal to the stationary loss probability for the
exponential distribution of job length.  Denote by $\tilde{p}_n$ the stationary loss probability for the system considered in Subsubsection~3.2. We assume that  $\alpha=0.5.$  
     
\vskip 25pt
{\bf Table 1.} Results of calculations
\vskip 10pt
{\small
\begin{tabular}{|c|c|c|c|c|c||c|c|c|c|c|c|}
\hline
No.&$n$&$\lambda$&$p_n^*$&$\hat{p}_n$&$\tilde{p}_n$&No.&$n$&$\lambda$&$p_n^*$&$\hat{p}_n$&$\tilde{p}_n$\\
\hline 
1&1&0.1&0.095&0.091&0.083&14&2&0.4&0.132&0.103&0.068\\
2&1&0.2&0.181&0.167&0.143&15&2&0.6&0.259&0.184&0.123\\
3&1&0.3&0.259&0.231&0.188&16&2&0.8&0.395&0.262&0.165\\
4&1&0.4&0.330&0.286&0.222&17&2&1&0.523&0.333&0.200\\
5&1&0.5&0.393&0.333&0.250&18&2&1.2&0.634&0.396&0.229\\
6&1&0.6&0.451&0.375&0.272&19&2&1.4&0.725&0.427&0.251\\
7&1&0.7&0.503&0.412&0.292&20&2&1.6&0.797&0.496&0.275\\
8&1&0.8&0.551&0.444&0.292&21&2&1.8&0.851&0.536&0.292\\
9&1&0.9&0.593&0.474&0.321&22&2&2&0.892&0.571&0.308\\
10&1&1&0.632&0.500&0.333&23&5&0.5&0.026&0.016&0.011\\
11&1&1.5&0.777&0.600&0.375&24&5&1&0.390&0.167&0.091\\
12&1&2&0.865&0.667&0.400&25&5&1.5&0.820&0.371&0.128\\
13&2&0.2&0.037&0.032&0.024&26&5&2&0.964&0.508&0.262\\
\hline 
\end{tabular}
}

\vskip 5pt

From the results of computations, we see that the loss probability (for the shortest remaining length rule) computed in accordance with (3) may be not decreasing function of $n$ for prescribed $\lambda$ and $b$ unlike that, obviously, the loss probability (for the related system) computed in accordance with  (6)--(8), (12) is a monotonically decreasing function of $n.$ If the arrival rate is sufficiently great, then the stationary probability that the number of jobs in the system is equal to $n$ is great. If the system is in this state, the jobs are served slowly. The service of a job in this state is interrupted with a great probability, and the served job is displaced. The length of the arriving job is greater than the remaining length of the interrupted job. For example, if $\lambda=2,$ $b=1,$ then $p_n^*$ is equal to 0.865, 0.892, 0.965 for $n$ equal to 1, 2, 5 respectively.  If $\lambda=1,$ then the value $p_n^*$ is equal to 0.635, 0.523, 0.385 for $n$ equal to 1, 2, 5 respectively. The loss probability is decreasing function of $n.$  If $\lambda=1.2,$ then the value $p_n^*$ is equal to 0.609, 0.634, 0.633 for 1, 2, 10 respectively. For $\lambda=1.6,$ we have  $p_n^*=0.796$ if $n=2,$ and, for a less value $\lambda=1.4,$ we have $p_n^*=0.848>0.796$ if $n=10.$

\section{Conclusion}

In this paper we  study a queueing system, in which there are no more than $n$ jobs at any time. If there are $i<n$ jobs in the system, then each job is served with rate 
$c_i,$ $i=1,\dots,n,$ and $c_1=1.$ The length of a job is the duration of this job service with rate 1. If there are $n$ jobs in the system and a new job arrives, then one of the jobs is lost. In accordance with one of the considered discipline, the job with the shortest remaining   
length is lost. Denote by $p_i(t)$ the probability that, at time $t,$ there $i$ jobs in the system, $i=1,\dots,n.$ Denote by $\hat{p}_i(t)$ the probability that, at time $t,$ there $i$ jobs in the corresponding unlimited processor sharing system, $i=1,2,\dots$ If there are 
$i< n$ jobs in the system, then each job is served with rate $c_i,$ $i=1,\dots,n-1.$ If there
are $i\ge n$ jobs in the system, then each job is served with rate $c_n.$

Assume that the initial state of the two systems is the same.   
It is proved at this paper that
$$p_i(t)=\hat{p}_i(t),\ i=1,\dots,n-1,$$ 
$$p_n(t)=\sum\limits_{i=n}^{\infty}\hat{p}_i(t)=1-\sum\limits_{i=0}^{n-1}\hat{p}_i(t).$$

In the case of a Poisson arrival process, stationary state probabilities of the unlimited processor sharing system do not depend on the service time distribution under the assumption that the expectation of this time is prescribed. Thus the
stationary probabilities are also invariant for the considering loss limited processor sharing system.
We have obtained explicit formulas for the stationary state probabilities and the loss probabilities for this case.

In the case of constant service time, the rule of the shortest remaining length job loss is equivalent to the FCFD rule. In the case of an arbitrary job length distribution, the value of the loss probability for the system with the rule of the shortest remaining length job 
loss is an upper bound for the loss probability in the corresponding system with an arbitrary rule of choosing a job to be lost.

The FCFD protocol minimizes the loss probability in the case of higher variance of the service time distribution.
  
We have found that the, in the case of FCFD rule and constant service length, 
stationary lost probability in a limited egalitarian processor sharing system can be not monotonically decreasing function of the maximum admissible number of jobs in the system for a fixed arrival rate unlike the system with the loss of arriving jobs, for which the loss probability is decreasing function. The stationary loss probability can be also not decreasing function of the maximum admissible number in the system for the shortest remaining length rule and an arbitrary job length distribution.

\section*{Acknowledgments}

The authors would like to thanks CNPq, Senai Cimatec,
Salvador BA, the Moscow Automobile and Road Construction
State Technical University (MADI)  and
Iecom, for the support and funding for the research.  
Also we would like to thanks our colleague Ivan A. Kuteynikov for  the calculations
presented in Table 1.

\bibliographystyle{unsrt}  


\end{document}